\documentclass{jpp}
\usepackage{graphicx}
\usepackage{epstopdf, epsfig}
\usepackage{hyperref}
\usepackage{amsmath}

\renewcommand{\vec}[1]{\boldsymbol{#1}}

\shorttitle{Particle motion in circularly polarized wave fields}
\shortauthor{C. Schreiner, R. Vainio and F. Spanier}

\title{Analytical treatment of particle motion in circularly polarized slab-mode wave fields}

\author{
  Cedric Schreiner\aff{1}
  \corresp{\email{mail@cschreiner.de}},
  Rami Vainio\aff{2}
  \and Felix Spanier\aff{3}
}

\affiliation{
  \aff{1}Max-Planck-Institute for Solar System Research, Justus-von-Liebig-Weg 3, DE-37077 G\"ottingen, Germany
  \aff{2}Department of Physics and Astronomy, University of Turku, FI-20014 Turku, Finland
  \aff{3}Centre for Space Research, North-West University, 2520 Potchefstroom, South Africa
}

\begin{document}

\maketitle

\begin{abstract}
Wave-particle interaction is a key process in particle diffusion in collisionless plasmas.
We look into the interaction of single plasma waves with individual particles and discuss under which circumstances this is a chaotic process, leading to diffusion.
We derive the equations of motion for a particle in the fields of a magnetostatic, circularly polarized, monochromatic wave and show that no chaotic particle motion can arise under such circumstances.
A novel and exact analytic solution for the equations is presented.
Additional plasma waves lead to a breakdown of the analytic solution and chaotic particle trajectories become possible.
We demonstrate this effect by considering a linearly polarized, monochromatic wave, which can be seen as the superposition of two circularly polarized waves.
Test particle simulations are provided to illustrate and expand our analytical considerations.
\end{abstract}

\keywords{plasma waves, plasma simulation}

\section{Introduction}
\label{sec:intro}

Most astrophysical plasmas are collision-free even on very large scales due to the extremely low particle densities.
The motion of charged particles is, therefore, not governed by collisions, but by interactions with the ubiquitous turbulent magnetic fields \citep{bell_1978,schlickeiser_1989}.
It is a common assumption that this interaction can be described in terms of wave-particle interaction.
This process is relevant in astro- and space physics, particle acceleration, plasma heating, or the physics of fusion devices. From the interaction of waves and particles all relevant transport parameters (i.e., Fokkker-Planck coefficients, mean free path) can be calculated in a typically very tedious procedure.

As wave-particle interaction is a fundamental process, the subject has been frequently revisited during the past decades, oftentimes focusing on relatively simple toy models to demonstrate basic features of more complex physical processes.
While this is generally not the case in real world plasmas, models may consider only a single or a limited number of individual plasma waves in order to investigate particle transport.
For the simple case of a single, circularly polarized electromagnetic wave propagating along a background magnetic field, the equations of motion of a charged particle are integrable and exact solutions can be found \citep{roberts_1964, arnold_1978, kong_2007}.
Furthermore it can be shown that particles can be trapped in potential wells created by the fields of the wave \citep{sudan_1971, schreiner_2017_a}.

However, this is only a special case of a much wider field of problems.
Obliquely propagating waves or waves propagating perpendicular to the background field, for instance, introduce stochastic behavior of the particles above a certain threshold amplitude of the wave \citep{palmadesso_1972, smith_1978, valvoglis_1984}.
Chaotic particle orbits may also be encountered as soon as a second wave comes into play.
Formerly trapped particles may be de-trapped by a second wave \citep{sakai_1972}, and the equations of motion are in general no longer integrable.
Depending on the phase space positions of the wave-particle resonances and the amplitudes of the waves, stochastic behavior may occur in limited regions of phase space \citep{murakami_1982, tran_1982, balakirev_1989, shklyar_2014}.
Besides these regions of chaotic particle motion individual islands exist in phase space, where different types of closed orbits can be found \citep{lehmann_2010}.

One might expect that at least a single, linearly polarized wave propagating along a static background field would present a simple enough problem to find exact solutions to the equations of motion of a charged particle.
This is, however, not the case \citep{bourdier_2001, bourdier_2005, bourdier_2009b}:
While the equations can be integrated if the linearly polarized wave propagates in an unmagnetized medium, this no longer holds for the magnetized case.

To further add to the discussion of wave-particle interaction and whether the equations of motion of a charged particle are integrable in a specific setup, we revisit the case of a magnetostatic, monochromatic, circularly polarized wave in a magnetized plasma.
We present a new approach to describing the motion of particles in the magnetic field of the wave and derive a solution to the equations of motion which, to our best knowledge, has not been considered elsewhere.
The calculations of our dynamical system are shown in detail and are easy to follow.
We then make use of numerical simulations to illustrate and expand our results.

Our work is differing from previous attempts, such as \citet{qian_2000} since this work focuses on the integrability of the system and not the resonances of particles and waves.
With a potential application in space physics in mind, we choose the magnetostatic case instead of a scenario including electromagnetic waves \citep[such as][]{balakirev_1989, kong_2007, bourdier_2009b, essen_2015}.
In space plasmas, particle transport is dominated by the interaction with Alfv\'en waves, which are often assumed to be magnetostatic in analytic models, due to their low frequency and constant velocity $v_\mathrm{A} \ll c$, which is considerably smaller than the speed of light $c$.

The article is organized as follows:
We present the basic idea of our model for particle transport in magnetostatic slab fluctuations in Sect.~\ref{sec:model}.
A novel set of dynamic variables is motivated in Sect.~\ref{sec:equations_of_motion}, where the equations of motion are also presented.
We then discuss the motion of particles in the magnetic field of a circularly polarized wave in Sect.~\ref{sec:circular_wave}.
With our new approach an exact solution to the equations of motion can be derived, which holds without any further assumptions.
We present phase space trajectories which show that particles move on closed orbits and calculate equilibrium positions.
In Sect.~\ref{sec:linear_wave} we then address the case of a linearly polarized wave.
Using the same mathematical approach, we demonstrate that no analytical solution can be found and that no equilibrium points exist in this case.
Numerical simulations of test particles propagating in a prescribed magnetic field setup are used to illustrate our previous findings in Sect.~\ref{sec:simulations}.
First we validate our simulations by reproducing the analytical results derived for the case of a monochromatic, circularly polarized wave.
As a next step we introduce a second circularly polarized wave with opposite helicity and varying amplitude, resulting in a monochromatic wave with general elliptic polarization.
We find chaotic particle motion, but also islands in phase space, where closed orbits seem to be possible.
Section~\ref{sec:conclusions} concludes the article with a brief summary and discussion.

\section{Model}
\label{sec:model}

We consider a transverse perturbation $\vec{\delta B}$ on top of a constant ($z$-directed) magnetic field $\vec{B_0}$:
\begin{equation}
  \vec{B} = \delta B_x (z) \vec{e}_x + \delta B_y (z) \vec{e}_y +  B_0  \vec{e}_z,
\end{equation}
where $\vec{e_x}$, $\vec{e_y}$, and $\vec{e_z}$ are the unit vectors in $x$-, $y$-, and $z$-direction.
This can be obtained from the vector potential
\begin{equation}
  \vec{A}(x,z) = \int_0^z \delta B_y(z') \, dz' \, \vec{e}_x + \left(B_0 x -\int_0^z \delta B_x(z') \, dz' \right)\vec{e}_y,
  \label{eq:vecpota}
\end{equation}
since
\begin{eqnarray}
  \nabla \times \vec{A} &=& \begin{vmatrix}
    \vec{e}_x & \vec{e}_y & \vec{e}_z\\
    \partial_x & \partial_y & \partial_z \\
    A_x & A_y & 0
  \end{vmatrix}
  =-\vec{e_x}\frac{\partial A_y}{\partial z}+\vec{e}_y \frac{\partial A_x}{\partial z} + \vec{e}_z \frac{\partial A_y}{\partial x}\\
  &=&\delta B_x (z) \vec{e}_x + \delta B_y(z) \vec{e}_y + B_0 \vec{e}_z.
\end{eqnarray}
It should be noted that the choice of the vector potential is not unique. Alternatively the potential $\tilde{\vec A}$ can be used:
\begin{equation}
  \tilde{\vec A}(y,z) = \left(-B_0 y + \int_0^z \delta B_y(z') \, dz'\right)\vec{e}_x - \int_0^z \delta B_x(z') \, dz' \, \vec{e}_y.
\end{equation}

The Hamiltonian for a particle with mass $m$ and charge $q$ in a magnetic vector potential can be written as
\begin{equation}
  H(\vec r, \vec P) = c\sqrt{(\vec P - q \vec A)^2+m^2 c^2},
\end{equation}
with the speed off light $c$ and the generalized momentum $\vec{P} = \gamma m \, \dot{\vec{x}} + q \, \vec{A}$.
For $\vec A$ the Hamiltonian does not depend on $y$  while for $\tilde{\vec A}$ the Hamiltonian does not depend on $x$. The Hamiltonian may be written as
\begin{equation}
  H(x,z,\vec P) = c\sqrt{(P_x - q A_x)^2 + (P_y-q A_y)^2+P_z^2+m^2 c^2}.
  \label{eq:orighamil}
\end{equation}
When using the vector potential $\vec A$ the coordinate $y$ becomes cyclic. Therefore
\begin{equation}
  P_y = p_y + q A_y = p_y + q\left(B_0 x - \int_0^z \delta B_x(z') dz'\right)
\end{equation}
is a constant of motion.
The Hamiltonian $H= \gamma m c^2$ does not depend on time explicitly and is thus a constant of motion.
Subsequently,  the particle momentum $p$ is also a constant of motion.

Next we define the coordinate $\vec{x_F} = (x_F, y_F,0)$ of the base of the field line which connects the $x$-$y$-plane to the coordinate $(x,y,z)$.
To obtain $x_F$ we rewrite $P_y$ using the field line equations.
The equation for a field line in the $x$-$z$-plane is
\begin{equation}
  \frac{dx}{B_x} = \frac{dz}{B_0} \Rightarrow x = x_F + \int_0^z \frac{\delta B_x(z')}{B_0}dz'.\label{eq:fieldline_x}
\end{equation}
We rephrase the above expression to obtain
\begin{equation}
  q\left( B_0 x -\int_0^z\delta B_x(z') dz' \right) = q B_0 x_F.
\end{equation}
Note that the left-hand side is equal to $q\,A_y$, as defined in Eq.~\eqref{eq:vecpota}.
Thus, we find the relation
\begin{equation}
  P_y = p_y + q B_0 x_F. \label{eq:pxfieldline}
\end{equation}
For symmetry reasons, we can assume that there must be a constant of motion related to $P_x$ as well, which can be used to obtain $y_F$:
\begin{equation}
  \frac{dy}{B_y} = \frac{dz}{B_0} \Rightarrow y = y_F + \int_0^z \frac{\delta B_y(z')}{B_0}dz'.\label{eq:fieldline_y}
\end{equation}
Using the magnetic vector potential $\tilde{\vec A}$, $x$ becomes the cyclic variable, making
\begin{equation}
  \tilde{P_x} = p_x + q \tilde{A}_x = p_x + q\left(-B_0 y + \int_0^z \delta B_y(z') \, dz'\right)
\end{equation}
a constant of motion. Relating $\tilde{P}_x$ to $P_x$ we can then write
\begin{eqnarray}
  \tilde{P}_x &=& p_x + q \tilde{A}_x = P_x + q(\tilde{A}_x - A_x)= P_x - q B_0 y\\
  &=& p_x -q B_0 y_F.
  \label{eq:pyfieldline}
\end{eqnarray}

The Hamiltonian with the vector potential $\vec A$ then has three constants of motion:
\begin{equation*}
  H(x,z,\vec P) = c\sqrt{(P_x - q A_x)^2 + (P_y - q A_y)^2 + P_z^2 + m^2 c^2}; \ P_y ; \ \tilde{P}_x = P_x-qB_0y.
\end{equation*}
Clearly, the latter two do not commute with each other and, thus, the system in general is not integrable. Thus, there is a prospect of chaotic orbits.

In our model we will assume that all particles are injected at $(x_0,y_0,z_0)=(0,0,0)$ at time $t=0$. Since we are discussing a pure slab model this does not change the generality. Their initial velocity is assumed to be $(v_{0x},v_{0y},v_{0z})$. Since $\vec A$ as well as $\tilde{\vec A}$ are 0 at the origin, we can conclude that
\begin{eqnarray}
  P_y (t=0)= p_y(t=0) &= \gamma m v_{0y},\\
  \tilde{P}_x (t=0) = p_x(t=0) &= \gamma m v_{0x}.
\end{eqnarray}
As discussed earlier, both quantities are constants of motion and therefore conserved at later times.

Before solving the problem of the interaction with magnetic irregularities $\vec{\delta B}$, we will discuss the solution of the unperturbed system ($\delta B = 0$). Using the original Hamiltonian \eqref{eq:orighamil} with the vector potential \eqref{eq:vecpota} it is easy to see that the solution is
\begin{eqnarray}
  x(t) &=&  \frac{1}{B_0 q} \left( \phantom{-}p_{0x} \sin\left(\frac{q B_0}{\gamma m} t \right) + p_{0y} \left( 1 - \cos\left(\frac{q B_0}{\gamma m} t \right) \right) \right),\\
  y(t) &=&  \frac{1}{B_0 q} \left( -p_{0x} \left( 1 - \cos\left(\frac{q B_0}{\gamma m} t \right) \right) + p_{0y} \sin\left(\frac{q B_0}{\gamma m} t \right) \right), \\
  z(t) &=&  \frac{p_{0z} t}{\gamma m}.
\end{eqnarray}
The straight-forward solution shows that particles gyrate around the gyrocenter at $(x,y) = (p_{0y}/(B_0 q),-p_{0x}/(B_0 q))$.
It should be noted that the particles' gyro-centers do not leave their original magnetic field line. However, the field line coordinate $(x_F,y_F)$ does not denote the field line of the particle's guiding center, but the field line currently connecting the $z=0$ plane with the actual particle.

\section{Equations of motion}
\label{sec:equations_of_motion}

From the Hamilton equations we see that
\begin{eqnarray}
  \dot x = v_x &=& \frac{\partial H}{\partial P_x} = \frac{P_x - q A_x}{\gamma m }, \\
  \dot y = v_y &=& \frac{\partial H}{\partial P_y} = \frac{P_y - q A_y}{\gamma m },\\
  \dot z = v_z &=& \frac{\partial H}{\partial P_z} = \frac{P_z}{\gamma m }.
\end{eqnarray}
We will now take three steps to derive a system with less variables:
\begin{enumerate}
  \item Since the total momentum is conserved, we may express the velocity in $z$-direction using the $x$- and $y$-components of the velocity vector.
  \item Instead of using the generalized momentum $P_x$ we will switch to the alternative formulation $\tilde{P}_x$.
  \item The magnetic vector potential will be expressed using the field line coordinates $(x_F, y_F)$ according to Eqs.~\eqref{eq:pxfieldline} and~\eqref{eq:pyfieldline}.
\end{enumerate}
The first two steps are trivial, the last step requires some caution: The introduction of the coordinates $(x_F, y_F)$ is first of all a mathematical construct. We have discussed earlier that the particle can change the field line which connects it to the $z=0$ plane without leaving the field line it is ``tied to'', i.e.~the field line along which the gyrocenter of the particle is traveling.

Applying these changes yields
\begin{eqnarray}
  \dot x = v_x &=& \frac{\tilde{P}_x + q B_0 y_F(y,z)}{\gamma m}, \\
  \dot y = v_y &=& \frac{P_y - q B_0 x_F(x,z)}{\gamma m},\\
  \dot z = v_z &=& \pm\sqrt{v^2-v_x^2(y,z)-v_y^2(x,z)}.
\end{eqnarray}
Using the initial conditions introduced in the model, we find
\begin{eqnarray}
  \dot x &=& v_x = v_{0x} + \Omega_0 y_F 
  ,  \label{eq:dotx}\\
  \dot y &=& v_y = v_{0y} - \Omega_0 x_F 
  ,\label{eq:doty}\\
  \dot z &=& v_z = \pm\sqrt{v^2-v_x^2(y,z)-v_y^2(x,z)},
\end{eqnarray}
where $\Omega_0 = q B_0 / \gamma m$ is the cyclotron frequency of the particle.

Since there is a bijective connection between the particle coordinates $(x,y,z)$ and the coordinates $(x_F,y_F,z)$, consisting of the field line coordinates $(x_F,y_F,0)$ and the $z$-coordinate of the particle, we may express the equations of motion also in these coordinates:
\begin{eqnarray}
  \dot z &=& \pm\sqrt{v^2-v_x^2(y,z)-v_y^2(x,z)}\nonumber\\
  &=& \pm \sqrt{v^2-(v_{0x}+\Omega_0 y_F)^2-(v_{0y}-\Omega_0 x_F)^2}.
\end{eqnarray}
The motion of $x$ and $y$ in field line coordinates can be derived from the time derivatives of Eqs.~\eqref{eq:fieldline_x} and~\eqref{eq:fieldline_y}:
\begin{eqnarray}
  \dot x &=& \dot{x}_F + \dot z \frac{\delta B_x}{B_0},\\
  \dot y &=& \dot{y}_F + \dot z \frac{\delta B_y}{B_0}.
\end{eqnarray}
These may be inserted into Eqs. \eqref{eq:dotx} and \eqref{eq:doty}:
\begin{eqnarray}
  \dot{x}_F &=& v_{0x} + \Omega_0 y_F - \dot{z}\frac{\delta B_x}{B_0},\\
  \dot{y}_F &=& v_{0y} - \Omega_0 x_F - \dot{z}\frac{\delta B_y}{B_0}.
\end{eqnarray}

In the next transformation step we will move to the coordinate set $(X_G, Y_G)$, which describes the position of the field line at $z=0$ relative to the initial guiding center position:
\begin{eqnarray}
  X_G &=& x_F - \frac{v_{0y}}{\Omega_0} = -\frac{v_y}{\Omega_0},\label{XG}\\
  Y_G &=& y_F + \frac{v_{0x}}{\Omega_0} = \frac{v_x}{\Omega_0}.\label{YG}
\end{eqnarray}

The equations of motion in this coordinate set are then
\begin{eqnarray}
  \dot{X}_G &=&  \phantom{-} \Omega_0 Y_G - \dot{z}\frac{\delta B_x}{B_0},\label{eq:dot_xg}\\
  \dot{Y}_G &=&  - \Omega_0 X_G - \dot{z}\frac{\delta B_y}{B_0},\label{eq:dot_yg}\\
  \dot{z} &=& \pm \sqrt{v^2-\Omega_0^2(X_G^2 + Y_G^2)}.\label{eq:dot_zg}
\end{eqnarray}
The coordinates with index $G$ describe the circular motion of the base of the field line. It is therefore advisable to use polar coordinates:
\begin{eqnarray}
  X_G &=& R_G \cos \varphi_G,\\
  Y_G &=& R_G \sin \varphi_G,
\end{eqnarray}
for which the following equations of motion are derived:
\begin{eqnarray}
  \dot{R}_G &=& \phantom{-}\dot{X}_G \cos \varphi_G + \dot{Y}_G \sin\varphi_G,\\
  R_G \dot{\varphi}_G &=& -\dot{X}_G \sin \varphi_G + \dot{Y}_G \cos\varphi_G.
\end{eqnarray}
Using Eqs.~(\ref{eq:dot_xg} - \ref{eq:dot_zg}) we can write:
\begin{eqnarray}
  \dot{R}_G &=& -\dot z \left(\frac{\delta B_x(z)}{B_0}\cos \varphi_G + \frac{\delta B_y(z)}{B_0}\sin \varphi_G \right),\\
  R \dot{\varphi}_G &=& \phantom{-} \dot z \left(\frac{\delta B_x(z)}{B_0}\sin \varphi_G - \frac{\delta B_y(z)}{B_0}\cos \varphi_G \right) -\Omega_0 R_G,\\
  \dot z &=& \pm \sqrt{v^2 - \Omega_0 R_G^2}.\label{eq:dot_z_R_G}
\end{eqnarray}
We solve the last equation for $R_G$ using $\dot z = v_z$ and obtain $R_G = \Omega_0^{-1} \sqrt{v^2 - v_z^2}$. The time derivative of $R_G$ then reads
\begin{equation}
  \dot{R}_G = -\frac{\dot{v}_z v_z}{\Omega_0 \sqrt{v^2-v_z^2}},
\end{equation}
and can be inserted into the equations of motion:
\begin{eqnarray}
  \dot{v}_z &=& \Omega_0 \sqrt{v^2-v_z^2}\left(\frac{\delta B_x(z)}{B_0}\cos \varphi_G + \frac{\delta B_y(z)}{B_0}\sin \varphi_G \right),\\
  \dot{\varphi}_G &=& -\Omega_0  + \Omega_0 \frac{v_z}{\sqrt{v^2-v_z^2}} \left(\frac{\delta B_x(z)}{B_0}\sin \varphi_G - \frac{\delta B_y(z)}{B_0}\cos \varphi_G \right),\\
  \dot z &=& v_z.
\end{eqnarray}
Using the definition of the pitch-angle cosine $\cos\theta = \mu = v_z / v$, we find the final set of equations:
\begin{eqnarray}
  \dot{\mu} &=& \Omega_0 \sqrt{1-\mu^2}\left(\frac{\delta B_x(z)}{B_0}\cos \varphi_G + \frac{\delta B_y(z)}{B_0}\sin \varphi_G \right),
  \label{eq:dot_mu}\\
  \dot{\varphi}_G &=& -\Omega_0  + \Omega_0 \frac{\mu}{\sqrt{1-\mu^2}} \left(\frac{\delta B_x(z)}{B_0}\sin \varphi_G - \frac{\delta B_y(z)}{B_0}\cos \varphi_G \right),
  \label{eq:dot_phiG}\\
  \dot z &=& v \mu.
  \label{eq:dot_z}
\end{eqnarray}

\section{Circularly polarized wave}
\label{sec:circular_wave}

So far we have only specified that the fluctuations are magnetostatic slab fluctuations. We will now focus on the circularly polarized case
\begin{equation}
  \vec{\delta B} = B_0 \epsilon \left( \cos \phi(z) \, \vec{e}_x - \sin \phi(z) \, \vec{e}_y \right),
  \label{eq:magcirc}
\end{equation}
with the relative amplitude $\epsilon = |\vec{\delta B}| / B_0$ of the plasma wave.
Here the equations of motion, Eqs.~(\ref{eq:dot_mu} - \ref{eq:dot_z}), reduce to
\begin{eqnarray}
  \dot \mu &=& \epsilon \Omega_0 \sqrt{1-\mu^2} \cos(\varphi_G + \phi(z)),\\
  \dot {\varphi}_G &=& - \Omega_0 + \epsilon \Omega_0\frac{\mu}{\sqrt{1-\mu^2}} \sin(\varphi_G + \phi(z))\label{eq:57},\\
  \dot z &=& v \mu \label{eq:58}.
\end{eqnarray}
These equations are exact.

The radius of motion of the particle mapped to the $z=0$ plane (using $X_G$ and $Y_G$ as defined in Eqs.~\eqref{XG} and~\eqref{YG}) is
\begin{equation}
  R_G = \frac{v}{\Omega_0}\sqrt{1-\mu^2},
\end{equation}
as we have derived from Eq.~\eqref{eq:dot_z_R_G}.
This expression is identical to the gyro-radius in the unperturbed field $B_0$.

\subsection{Monochromatic wave}

In the first step we have described the polarization of the wave, but the phase relation $\phi(z)$ has not been specified. The simplest case is the monochromatic wave with $\phi(z) = k z $. The phase between wave and particle is given by
\begin{equation}
  \psi = \varphi_G + k z.\label{eq:phase_angle}
\end{equation}

Using Eqs.~\eqref{eq:58} and \eqref{eq:phase_angle} its time derivative is obtained as
\begin{equation}
  \dot \psi = \dot{\varphi}_G + k \mu v,
\end{equation}
and substituting $\dot{\varphi}_G$ from Eq.~\eqref{eq:57} gives us a set of two autonomous equations describing particle motion:
\begin{eqnarray}
  \dot \mu &=& \epsilon \Omega_0 \sqrt{1-\mu^2 } \cos\psi\label{eq:finalmu},\\
  \dot \psi &=& - \Omega_0 + k v \mu +\epsilon \Omega_0 \frac{\mu}{\sqrt{1-\mu^2}} \sin \psi\label{eq:finalpsi}.
\end{eqnarray}
This already shows that there are no chaotic solutions to the problem.
By dividing one of the above equations by the other, a first order ordinary differential equation for one of the variables (either $\psi$ or $\mu$) as a function of the second variable can be obtained \citep[see e.g.][Chapter~7]{prelle_1983, teschl_2012}.
The resulting equation can be solved for a family of curves and one constant of integration, which then fixes the correct path of a particle in phase space.
This can be done for all systems which depend on only two variables, while a minimum of three variables is required for chaotic behavior.

For the case of Eqs.~\eqref{eq:finalmu} and~\eqref{eq:finalpsi} we obtain a family of curves in the $(\psi,\mu)$ plane:
\begin{eqnarray}
  \frac{d\psi}{d\mu} &=& \frac{-\Omega_0 + k v\mu + \Omega_0 \frac{\mu}{\sqrt{1-\mu^2}}\epsilon \sin\psi}{\Omega_0 \sqrt{1-\mu^2}\epsilon \cos\psi},
  \\
  \frac{d\epsilon\sin\psi}{d\mu}= \epsilon \cos \psi \frac{d\psi}{d\mu} &=& \frac{-\Omega_0 + k v\mu + \Omega_0 \frac{\mu}{\sqrt{1-\mu^2}}\epsilon \sin\psi}{\Omega_0 \sqrt{1-\mu^2}}.
\end{eqnarray}

In order to find a relation between $\mu$ and $\psi$, which allows to draw the particles' paths in $(\psi,\mu)$ coordinates, we define a function $F(\mu)=\epsilon \sin\psi$. We can show the following relation:
\begin{eqnarray}
  \frac{dF}{d\mu} - \frac{\mu}{1-\mu^2} F &=& \frac{kv\mu- \Omega_0}{\Omega_0\sqrt{1-\mu^2}}\label{eq:gl64},\\
  \frac{1}{\sqrt{1-\mu^2}}\frac{d}{d\mu}(\sqrt{1-\mu^2}F) &=& \frac{kv\mu- \Omega_0}{\Omega_0\sqrt{1-\mu^2}},\\
  \frac{d}{d\mu}(\sqrt{1-\mu^2}F) &=& \frac{kv\mu- \Omega_0}{\Omega_0},\\
  \sqrt{1-\mu^2}F &=& \frac{(kv\mu- \Omega_0)^2-C}{2 k v \Omega_0},\\
  \epsilon \sin \psi = F &=& \frac{(kv\mu- \Omega_0)^2-C}{2 k v \Omega_0\sqrt{1-\mu^2}}.\label{eq:xx}
\end{eqnarray}
The problem has been reduced to one integration and is now identical to the motion along $\mu$ in a given potential.
We use $\epsilon \cos\psi = \sqrt{\epsilon^2 - \epsilon^2 \sin^2\psi}$ and are thus able to reformulate Eq.~\eqref{eq:finalmu}.
We insert the expression on the right-hand side of Eq.~\eqref{eq:xx} and eliminate $\psi$ from the equation.
After separating the variables $\mu$ and $t$ we are finally able to write the formal integral:
\begin{equation}
  t = \pm \int_{\mu_0}^{\mu(t)} \frac{2 \, k \, v \, d\mu'}{\sqrt{4\epsilon^2\Omega_0^2 k^2 v^2(1-\mu'^2) - [(k v \mu' - \Omega_0)^2 -C]^2}}.
\end{equation}

\begin{figure}
  \centering
  \includegraphics[width=0.8\linewidth]{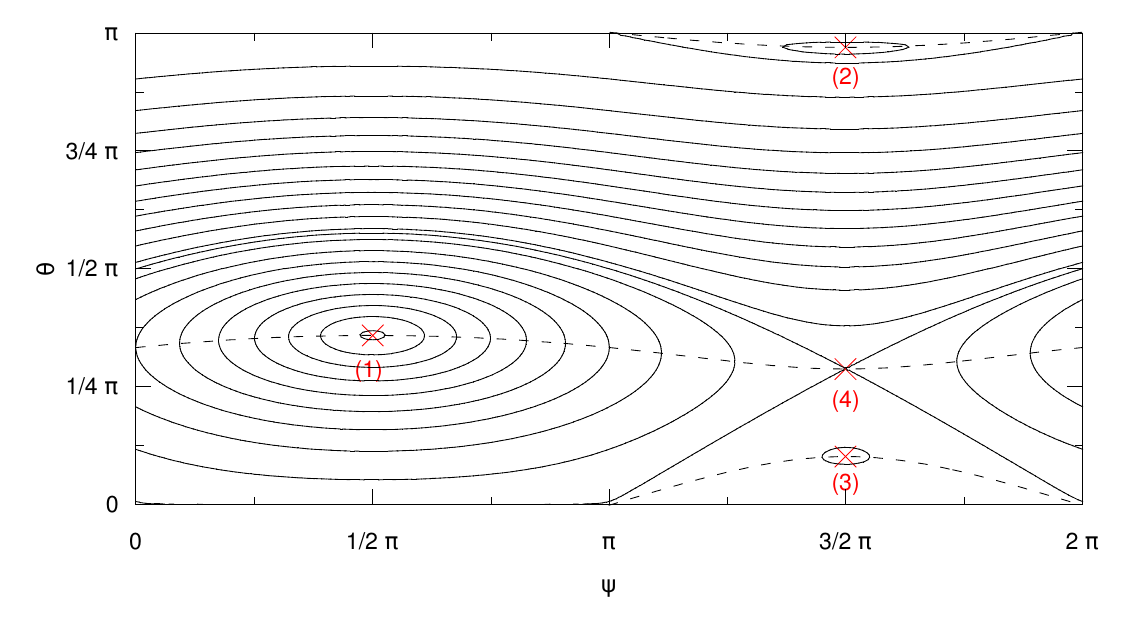}
  \caption{
  Phase space trajectories (solid curves) for a circularly polarized wave with $k = 2 \Omega_0 /v$ and $\epsilon = 0.3$.
  The dashed curves give the track of $d\psi = 0$  along the trajectories.
  Four equilibrium points -- three stable ones (labeled 1,\,2,\,3) and one unstable (4) -- can be seen at the intersection of the dashed curves and lines $\psi=\frac{1}{2}\upi$ and $\psi=\frac{3}{2}\upi$.
  }
  \label{fig:1}
\end{figure}

We rearrange Eq. \eqref{eq:xx} to find an expression for the constant of integration, $C$:
\begin{equation}
  C = (kv\mu-\Omega_0)^2 - 2\Omega_0 k v \sqrt{1-\mu^2}\epsilon \sin \psi,
  \label{eq:const_integration}
\end{equation}
which is a constant of motion. Thus, the value of $C$ can be calculated at $t=0$. The contours for $C$ give the orbits in the $(\psi,\mu)$ space, as shown in Fig. \ref{fig:1}. For the trajectories we may easily find the position of $d\psi/d\mu = 0$ along the curves:
\begin{eqnarray}
  0 &=& dC = \frac{\partial C}{\partial \mu}d\mu + \frac{\partial C}{\partial \psi}d\psi,\\
  0 = \frac{d\psi}{d\mu} &=& - \frac{\frac{\partial C}{\partial\mu}}{\frac{\partial C}{\partial\psi}}\Leftrightarrow 0 =\frac{\partial C}{\partial\mu} = 2kv(kv\mu-\Omega_0)+2\Omega_0kv\frac{\mu}{\sqrt{1-\mu^2}}\epsilon\sin\psi,\\
  0 &=& \left( \frac{kv\mu}{\Omega_0}-1\right)\sqrt{1-\mu^2}+\epsilon \mu\sin \psi.
\end{eqnarray}
This relation describes the curves connecting the turning points of the closed contours, as shown by the dashed lines in Fig.~\ref{fig:1}.
At the intersections with the dashed lines the particle trajectories (represented by the solid lines) are vertical, i.e.~$d\psi = 0$.

So far we have discussed particle orbits in $\psi$-$\mu$-phase space, where $\mu$ is the pitch-angle cosine in the unperturbed magnetic field $\vec{B}_0$.
It should be kept in mind that the trajectories have different characteristic when the actual pitch-angle $\alpha$ -- with \mbox{$\cos \alpha = \vec{v} \cdot \vec{B} / (|\vec{v}| \, |\vec{B}|)$} -- is considered.
We discuss the relation of $\cos \alpha$ and $\mu$ in appendix~\ref{app:actual_pitch_angle}.

\subsection{Equilibrium points}

\begin{figure}
  \centering
  \includegraphics[width=0.8\linewidth]{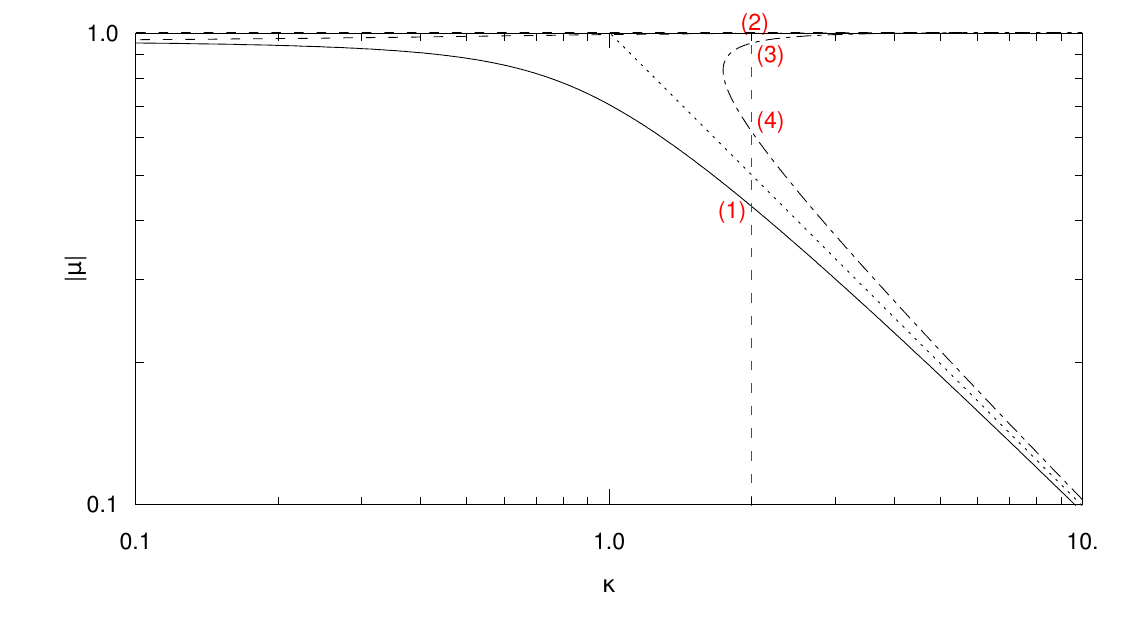}
  \caption{
  Equilibrium points for $\mu=\mu(\kappa)$ for $\epsilon=0.3$.
  The solid curve gives the positive root at $\psi=\frac{1}{2}\upi$.
  The dashed and dash-dotted curves give the roots at $\psi=\frac{3}{2}\upi$ for $\mu<0$ and $\mu>0$, respectively.
  The dotted line is the limit $\mu=1/\kappa$ at $\epsilon=0$.
  The red dashed line marks $\kappa=2$ and the labels at the intersections with the black lines correspond to the equilibrium points in Fig.~\ref{fig:1}.
  }
  \label{fig:2}
\end{figure}

The equations of motion \eqref{eq:finalmu} and \eqref{eq:finalpsi} can be rewritten as
\begin{eqnarray}
  \dot \mu &=& \Omega_0\epsilon\sqrt{1-\mu^2}\cos\psi \equiv \Omega_0 f(\mu,\psi),
  \label{eq:dot_mu_equilibrium}
  \\
  \dot \psi &=& \Omega_0\left( -1+\kappa\mu+\epsilon\frac{\mu}{\sqrt{1-\mu^2}}\sin\psi \right)\equiv \Omega_0 g(\mu,\psi),
  \label{eq:dot_psi_equilibrium}
\end{eqnarray}
with $\kappa=kv/\Omega_0$
\footnote{In the limit of vanishing amplitude $\epsilon \rightarrow 0$ (an assumption that is sometimes made when deriving transport theory) the parameter $\kappa$ obtains a physical meaning as the inverse of the cosine $\mu_\mathrm{res}$ of the resonant pitch-angle.
Thus, for $\mu \rightarrow \mu_\mathrm{res}$ we find $\dot\psi = 0$ according to Eq.~\eqref{eq:dot_psi_equilibrium}.
}.
The equilibrium is defined by $\dot\mu=0 $ and $\dot\psi=0$. The first equation yields the requirement $\cos\psi=0$ and we can conclude
\begin{equation}
  \kappa \mu = 1- \epsilon\frac{\mu\sin\psi}{\sqrt{1-\mu^2}},\label{eq:y}
\end{equation}
i.e.,
\begin{eqnarray}
  \kappa \mu = \begin{cases}
    1-\epsilon\frac{\mu}{\sqrt{1-\mu^2}} & \text{for\ } \psi=\frac{\upi}{2},\\
    1+\epsilon\frac{\mu}{\sqrt{1-\mu^2}} & \text{for\ } \psi=\frac{3\upi}{2}.
  \end{cases}
\end{eqnarray}
The quartic equation
\begin{equation}
  (\kappa\mu-1)^2(1-\mu^2) = \epsilon^2\mu^2\label{equilibrium_points}
\end{equation}
describes the position of the equilibria. Only solutions $-1\leq\mu\leq 1$ are physical. There are either two or four equilibrium points for $\mu$ depending on the values of $\kappa$ and $\epsilon$. We plot the roots of Eq.~\eqref{equilibrium_points} in Fig.~\ref{fig:2} for $\epsilon=0.3$. The four equilibrium points for $\kappa=2$ are also visible in Fig.~\ref{fig:1}.

The stability of the equilibrium points is determined by the eigenvalues of
\begin{eqnarray}
  \mathsfbi{S} =\begin{pmatrix}
    \frac{\partial f}{\partial \mu} & \frac{\partial f}{\partial \psi} \\
    \frac{\partial g}{\partial \mu} & \frac{\partial g}{\partial \psi}
  \end{pmatrix}
  = \begin{pmatrix}
    -\epsilon\frac{\mu}{\sqrt{1-\mu^2}}\cos \psi & -\epsilon\sqrt{1-\mu^2}\sin\psi\\
    \kappa + \frac{\epsilon\sin\psi}{(1-\mu^2)^{3/2}} & \epsilon\frac{\mu}{\sqrt{1-\mu^2}}\cos \psi
  \end{pmatrix}.
\end{eqnarray}
We employ the condition $\cos\psi = 0$, which we have obtained from combining Eq.~\eqref{eq:dot_mu_equilibrium} with the requirement $\dot\mu=0$, to find the eigenvalues $\lambda$:
\begin{eqnarray}
  0 &=& \begin{vmatrix}
    -\lambda & -\epsilon\sqrt{1-\mu^2}\sin\psi \\
    \kappa + \frac{\epsilon\sin\psi}{(1-\mu^2)^{3/2}} & -\lambda
  \end{vmatrix}\\
  &=& \lambda^2 + \epsilon\sqrt{1-\mu^2}\sin\psi\left(\kappa + \frac{\epsilon\sin\psi}{(1-\mu^2)^{3/2}} \right),\\
  \lambda^2 &=& -\epsilon\sqrt{1-\mu^2}\sin\psi\left( \kappa + \frac{\epsilon\sin\psi}{(1-\mu^2)^{3/2}} \right)\\
  &=& -\epsilon\sqrt{1-\mu^2}\sin\psi\left( \frac{\kappa\mu}{\mu} + \frac{\epsilon\sin\psi}{(1-\mu^2)^{3/2}} \right)
\end{eqnarray}
In the next step we replace $\kappa\mu$ according to Eq.~\eqref{eq:y} and then make use of $\sin^2\psi=1$, since $\sin\psi=\pm1$:
\begin{eqnarray}
  \lambda^2 &=& -\epsilon\sqrt{1-\mu^2}\sin\psi\left( \frac{1}{\mu}\left(1-\mu\frac{\epsilon\sin\psi}{\sqrt{1-\mu^2}} \right) + \frac{\epsilon\sin\psi}{(1-\mu^2)^{3/2}} \right)\\
  &=& -\epsilon\sqrt{1-\mu^2}\left[ \frac{\sin\psi}{\mu} + \frac{\mu^2\epsilon}{(1-\mu^2)^{3/2}} \right].
\end{eqnarray}
We now have to consider the two cases $\sin\psi = \pm 1$.
For $\sin\psi=+1$ we find $\lambda^2 <0$ (marginally stable point) for all $\mu>0$ and
\begin{equation}
  \mu < - \frac{1}{\sqrt{1+\epsilon^{2/3}}}.
\end{equation}
Since the only root for $\sin\psi =+1$ has $\mu>0$ it is marginally stable (see label '(1)' in Figs.~\ref{fig:1} and~\ref{fig:2}).

For $\sin\psi=-1$ we obtain $\lambda^2<0$ for all $\mu<0$ and
\begin{equation}
  \mu > \frac{1}{\sqrt{1+\epsilon^{2/3}}}.
\end{equation}
We note that
\begin{equation}
  \kappa = \frac{1}{\mu}-\epsilon\frac{\sin\psi}{\sqrt{1-\mu^2}},
\end{equation}
so
\begin{eqnarray}
  \frac{\partial \kappa}{\partial\mu} =
  -\frac{1}{\mu^2}-\epsilon\frac{\mu\sin\psi}{(1-\mu^2)^{3/2}}
  = -\frac{1}{\mu} \left( \frac{1}{\mu} + \frac{\mu^2\epsilon\sin\psi}{(1-\mu^2)^{3/2}} \right)
\end{eqnarray}
and
\begin{equation}
  \lambda^2 = \epsilon \mu \sin\psi\sqrt{1-\mu^2} \frac{\partial \kappa}{\partial \mu}.
\end{equation}
This yields $\lambda^2 = 0 \Leftrightarrow \partial \kappa/\partial \mu = 0$.
Thus, the negative root is always stable (see label '(2)' in Figs.~\ref{fig:1} and~\ref{fig:2}) and the turning point $\partial|\mu|/\partial \kappa \to \infty$ at
\begin{equation}
  \mu = \frac{1}{\sqrt{1+\epsilon^{2/3}}}, \ \kappa =(1+\epsilon^{2/3})^{3/2}
\end{equation}
of the equilibrium point marked with the dot-dashed curve in Fig.~\ref{fig:2} denotes the loss of stability, when going toward lower values of $\mu$. The point is (marginally) stable at
\begin{equation}
  \mu > \frac{1}{\sqrt{1+\epsilon^{2/3}}}
\end{equation}
and unstable (a saddle) below (see labels '(3)' and '(4)', respectively, in Figs.~\ref{fig:1} and~\ref{fig:2}).

\section{Linear polarization}
\label{sec:linear_wave}

\subsection{Analytical approach}
\label{sec:linear_analytic}

For the case of linear polarization we may start by replacing Eq.~\eqref{eq:magcirc} with the following polarization:
\begin{equation}
  \vec{\delta B} = B_0 \epsilon \cos \phi(z) \vec{e}_x.
\end{equation}
According to Eqs.~(\ref{eq:dot_mu} - \ref{eq:dot_z}), this immediately yields the following equations:
\begin{eqnarray}
  \dot \mu &=& \Omega_0 \sqrt{1-\mu^2}\epsilon \cos \varphi_G \cos\phi(z),\\
  \dot{\varphi}_G &=& -\Omega_0 + \epsilon \Omega_0 \frac{\mu}{\sqrt{1-\mu^2}} \sin\varphi_G \cos\phi(z),\\
  \dot{z} &=& v \mu.
\end{eqnarray}
Using trigonometric identities the first and second equation are easily transformed into:
\begin{eqnarray}
  \dot \mu &=& \Omega_0 \sqrt{1-\mu^2} \epsilon \frac{1}{2}\Big( \cos\big(\varphi_G + \phi(z)\big) + \cos\big(\varphi_G-\phi(z)\big) \Big),\\
  \dot{\varphi}_G &=& -\Omega_0 + \frac{\epsilon \Omega_0}{2} \frac{\mu}{\sqrt{1-\mu^2}}\Big( \sin\big(\varphi_G + \phi(z)\big) + \sin\big(\varphi_G-\phi(z)\big) \Big).
\end{eqnarray}
These results are as expected and identical to the case of two waves with the same wave number and opposite magnetic helicity, $\sigma=\pm 1$.

Using the phase relation
\begin{equation}
  \phi(z) = k z,
\end{equation}
we find the equations:
\begin{eqnarray}
  \dot \mu &=& \Omega_0 \sqrt{1-\mu^2} \epsilon \frac{1}{2}\Big( \cos\big(\varphi_G + k z\big) + \cos\big(\varphi_G-k z\big) \Big),\\
  \dot{\varphi}_G &=& -\Omega_0 + \frac{\epsilon \Omega_0}{2} \frac{\mu}{\sqrt{1-\mu^2}}\Big( \sin\big(\varphi_G + k z\big) + \sin\big(\varphi_G- k z\big) \Big),\\
  \dot z &=& v \mu.
\end{eqnarray}
Following the previous calculations for circularly polarized waves, we define:
\begin{equation}
  \psi^\pm = \varphi_G \pm k z.
  \label{eq:psipm}
\end{equation}
Using these abbreviations we can calculate the following:
\begin{eqnarray}
  \dot \mu &=& \Omega_0 \sqrt{1-\mu^2} \frac{\epsilon}{2}\left( \cos\psi^+ + \cos\psi^-\right),\\
  \dot{\psi}^+ &=& -\Omega_0 + k v \mu + \frac{\epsilon \Omega_0}{2} \frac{\mu}{\sqrt{1-\mu^2}}\big( \sin\psi^+ + \sin\psi^- \big),\\
  \dot{\psi}^- &=& -\Omega_0 - k v \mu + \frac{\epsilon \Omega_0}{2} \frac{\mu}{\sqrt{1-\mu^2}}\big( \sin\psi^+ + \sin\psi^- \big).
\end{eqnarray}
These equations are a special case of the general expression for two waves, where the magnetic amplitudes are $\epsilon^+ = \epsilon^- = \epsilon/2$ and the wave numbers $k^+=-k^-=k$.

One is tempted to follow the approach used for circular polarization. There, a function $F(\mu)=\epsilon \sin\psi$ had been defined to provide a relation between $\psi$ and $\mu$. In the case of the circularly polarized wave this leads to Eq.~\eqref{eq:gl64}, which is an equation for $\psi$ only depending on $\mu$. Unfortunately, both $\psi^\pm$ depend on the other $\psi$. This makes the definition $F^\pm(\mu) = \epsilon \sin \psi^\pm$ impractical.

We have also tried $F(\mu) = \epsilon(\sin \psi^+ + \sin \psi^-)$, but we find that:
\begin{eqnarray}
  \frac{dF}{d\mu} &=& \epsilon\left(\frac{d \psi^+}{d\mu}\cos\psi^+ + \frac{d \psi^-}{d\mu}\cos\psi^-\right)\\
  &=& \frac{-2}{\sqrt{1-\mu^2}} + \frac{\epsilon \mu}{1-\mu^2} \big( \sin\psi^+ + \sin\psi^- \big) + \frac{2 k v \mu}{\Omega_0 \sqrt{1-\mu^2}} \frac{\cos\psi^+ - \cos\psi^-}{\cos\psi^+ + \cos\psi^-}.
\end{eqnarray}
Unfortunately, no combination of $\psi^\pm$ and $\sin \psi^\pm$ decouples $\psi^\pm$ from $\mu$.

We may -- on the other hand -- still try to find stationary points: $d\psi^\pm/d t = 0$ and $d\mu/d t = 0$:

\begin{eqnarray}
  \dot\mu &=& 0 = \Omega_0 \sqrt{1-\mu^2}\frac{\epsilon}{2}\big(\cos \psi^+ + \cos \psi^-\big),\\
  \dot\psi^\pm  &=& 0 = -\Omega_0 \pm k v \mu + \frac{\epsilon}{2} \frac{\mu}{\sqrt{1-\mu^2}}\big(\sin\psi^+ + \sin\psi^-\big).
\end{eqnarray}
The two stability conditions for $\psi^\pm$ can never be fulfilled simultaneously as can easily be seen when adding and subtracting them.
Thus, there is no stationary point in the $(\mu, \psi^+, \psi^-)$ phase space.

This result can also be illustrated by looking at the definition of $\psi^\pm$ in Eq.~\eqref{eq:psipm}.
For an equilibrium point we define the change $\Delta\psi^+$ of the relative phase between a particle and the first wave as
\begin{equation}
  \Delta\psi^+ = \Delta(\varphi_G + k \, z)=\Delta\varphi_G + k \, \Delta z = 0.
  \label{eq:condition_plus}
\end{equation}
At the same time we require $\Delta\psi^- = 0$ for the relative phase between the particle and the second wave.
However, this cannot be achieved due to Eq.~\eqref{eq:condition_plus}, since
\begin{equation}
  \Delta\psi^- = \Delta(\varphi_G - k \, z)=\Delta\varphi_G - k \, \Delta z=2 \, \Delta\varphi_G.
  \label{eq:condition_minus}
\end{equation}

These considerations show that a particle in resonance with one wave (i.e.~$\Delta\psi^+ = 0$) cannot be in resonance with the other wave at the same time.
Formally it is possible to satisfy both $\Delta\psi^+ = 0$ and $\Delta\psi^- = 0$, but according to Eq.~\eqref{eq:condition_minus} this requires $\Delta\varphi_G = 0$ -- a condition which implies that the particle is not gyrating since $\varphi_G$ is constant.

\subsection{Poincar\'e sections}
\label{sec:poincare}

We have shown that no stationary points in three-dimensional phase space can exist when a particle is propagating in the fields of a linearly polarized wave.
However, we might still be able to derive some more information about the behavior of the particles.
In Sect.~\ref{sec:simulations} we will make use of Poincar\'e sections to analyze particle motion.
A Poincar\'e section can be defined as a $(n-1)$-dimensional cut through an otherwise $n$-dimensional phase space, which then allows to reveal certain characteristics of a dynamical system.
Exact mathematical definitions can be found in standard textbooks, such as e.g.~\citet[][Chapters~6 and~12]{teschl_2012}.

By means of the Poincar\'e section a continuous dynamical system is transformed into a discrete system.
For the case of particle trajectories this means that the continuous path of a particle in phase space is reduced to a set of discrete points, marking the crossings of the Poincar\'e section.
Thus, periodic or quasi-periodic trajectories in $n$-dimensional space can be revealed, since they are represented by either a single point or an accumulation of nearby points in the $(n-1)$-dimensional section.
Such analysis is a typical method used to study chaotic systems and has been applied to the problem of charged particle transport before \citep[e.g.][]{murakami_1982, bouquet_1998, bourdier_2009b, lehmann_2010}.

We define our Poincar\'e sections by the condition $\psi^- = \psi^+ + n 2\upi$, $n \in \{0, 1, 2, \dots\}$.
In order to find the position of a particle trajectory within the Poincar\'e section, one has to integrate over the particle's orbit, which requires knowledge of the actual solution of the equation of motion.
For the case of periodic solutions the Poincar\'e section can show equilibria even though there is no stationary point in the full three-dimensional phase space.
These periodic (or quasi-periodic) solutions are marked by accumulations of points (or recurring patterns) in the Poincar\'e section.

\section{Magnetostatic particle simulations}
\label{sec:simulations}

We perform numerical test particle simulations to expand our results beyond the case described by the analytical model.
As a first step, however, we reproduce the analytical result for the case of a single, circular wave to validate our numerical approach.
The magnetostatic test particle code first presented by \citet{schreiner_2017_a} is used.
This code initializes a prescribed, static magnetic field and then uses the particle pusher of the Particle-in-Cell code ACRONYM \citep{kilian_2011} to propagate the test particles.

\subsection{Circular polarization}
\label{sec:simulations_one}

The test particle code allows to initialize a background magnetic field $B_0$ along $z$-direction together with the static magnetic field $\vec{\delta B}(z)$ of a wave with left-handed magnetic helicity, as described by Eq.~\eqref{eq:magcirc} for $k > 0$.
Test particles (in this case: protons) are injected as a mono-energetic population at the origin of the coordinate system, i.e.~$(x=0, \, y=0, \, z=0)$, in accordance with the assumptions made in Sect.~\ref{sec:model}.
The orientation of the velocity vectors of these particles is chosen at random, so that the particles spread out into different directions once the simulation commences.

In certain output intervals the position and velocity of each test particle are saved to disk.
This allows to track the individual particles and to compute their pitch-angles $\theta$ (or the corresponding cosines $\mu$) and the phase $\psi$ according to Eq.~\eqref{eq:phase_angle}.
Using Eqs.~\eqref{XG} and~\eqref{YG}, as well as the definition of the pitch-angle in the unperturbed case the two variables can be expressed by:
\begin{eqnarray}
  \psi &=& \arccos\left(\frac{-v_y}{\sqrt{v_x^2 + v_y^2}} + k z\right),\label{simulation_psi}\\
  \theta &=& \arccos\left(\frac{v_z}{v}\right).\label{simulation_theta}
\end{eqnarray}
The required parameters can be easily extracted from the simulation output.

\begin{figure}
  \centering
  \includegraphics[width=0.9\linewidth]{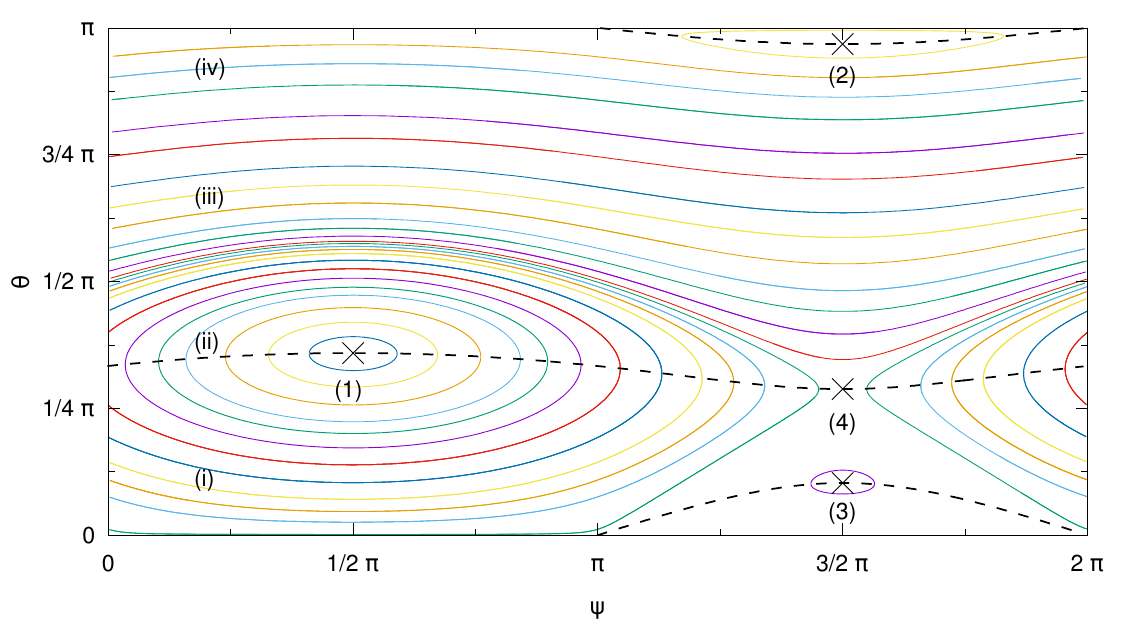}
  \caption{
    Phase space trajectories of test particles in a magnetostatic simulation.
    Each colored line represents the motion of a single particle.
    The simulation employs a single, circularly polarized wave with $k=2 \Omega_0 / v$ and $\epsilon = 0.3$.
    The particles follow the expected trajectories, as predicted by the analytic model (see Fig.~\ref{fig:1}).
    For verification we show the turning points $d\psi/d\mu = 0$ (black dashed lines) and positions of the equilibria (labeled with Arabic numbers), as discussed in Sect.~\ref{sec:circular_wave}.
    The four orange curves (labeled with Roman numbers) are discussed in Appendix~\ref{app:num_error}.
  }
  \label{fig:3}
\end{figure}

The simulation uses the same values $\epsilon=0.3$ and $\kappa=2$ as previously chosen for Fig.~\ref{fig:1}.
The three-dimensional simulation box with periodic boundary conditions consists of $128 \times 128 \times 2048$ grid cells in $x$-, $y$- and $z$-direction, which yields a high resolution of the field fluctuations along the $z$-direction caused by the plasma wave.
Perpendicular to $\vec{B}_0$ the spatial resolution is not important, since the field configuration does not change along the $x$- or $y$-axis.
The chosen extent of the simulation box allows to recover the gyro motion of the particles in space in order to obtain some more detailed information about the particles, which can be used for diagnostic purposes.
However, since Eqs.~\eqref{simulation_psi} and~\eqref{simulation_theta} only require information about the particles' velocity vectors and $z$-coordinates, a one-dimensional simulation setup would also be sufficient.

Phase space trajectories of test particles are shown in Fig.~\ref{fig:3}.
As can be seen, the magnetostatic test particle simulation is able to accurately reproduce the analytical results (see Appendix~\ref{app:num_error} for a discussion of numerical accuracy).
All particles follow closed orbits in phase space and the positions of the equilibrium points are recovered as expected.

\subsection{Other polarizations}
\label{sec:simulations_two}

Next we investigate the behavior of particles in the presence of two plasma waves.
We consider four setups (S1 through S4) which each include two circularly polarized waves with opposite magnetic helicity.
The first wave has $k^+>0$ and employs $\epsilon^+ = 0.3$ and $\kappa^+ = 2$, as specified in the previous section.
We choose $k^-=-k^+<0$ and thus $\kappa^-=-2$.
The amplitude $\epsilon^-$ of the second wave is varied in the different setups as shown in table~\ref{tab:amplitude}.
Depending on the the ratio $\epsilon^-/\epsilon^+$ the polarization of the superposition of both waves changes from circular to linear.

\begin{table}
  \centering
  \begin{tabular}{r c c c c}
      \textbf{setup} & S1 & S2 & S3 & S4\\
      $\boldsymbol{\epsilon^- / \epsilon^+}$ & 0.00 & 0.25 & 0.50 & 1.00\\
      \textbf{polarization} & circular & elliptic & elliptic & linear\\
  \end{tabular}
  \caption{Ratio $\epsilon^-/\epsilon^+$ of the amplitudes of the forward ($+$) and backward ($-$) propagating waves and the resulting polarization of their superposition.}
  \label{tab:amplitude}
\end{table}

\begin{figure}
  \centering
  \includegraphics[width=0.9\linewidth]{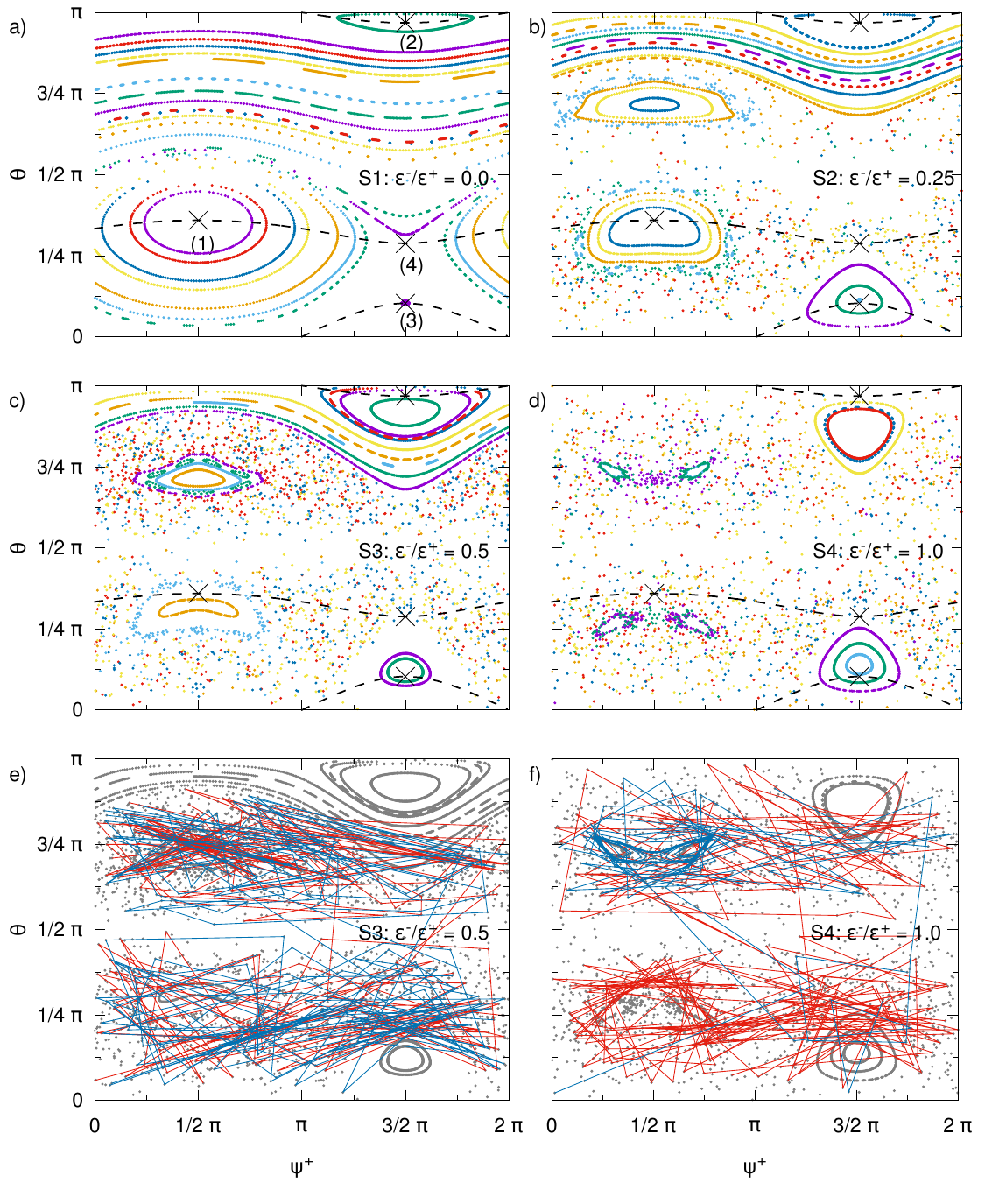}
  \caption{
    Colored dots represent phase space coordinates of individual particles on the $\psi^+$-$\theta$-plane defined by a Poincar\'e section.
    The black dashed lines and black crosses mark the turning points $d\psi/d\mu = 0$ and the equilibrium points (labeled 1 through 4 in panel a) for the case of a single circularly polarized wave.
    The different panels refer to the four setups S1, S2, S3, and S4 as specified by the label in each panel.
    Each label also includes the ratio $\epsilon^- / \epsilon^+$ of the amplitudes of the forward ($+$) and backward ($-$) propagating waves, as described in the text and in table~\ref{tab:amplitude}.
    The regions in which closed trajectories are possible shrink with increasing amplitude of the second plasma wave (panels a through d).
    In between these ordered regions chaotic behavior of the particles can be observed.
    For illustration, the dots representing the phase space position of two particles (red and blue) are connected in panels e and f.
    This gives an impression of particle motion in the Poincar\'e section and the chaotic orbits these particles are on.
    Note that lines crossing $\psi^+ = 0$ (or $\psi^+ = 2 \upi$) are not shown.
  }
  \label{fig:4}
\end{figure}

For graphical analysis, it is worthwhile to find a suitable representation of the system employing only two variables.
We therefore introduce
\begin{equation}
  \Delta\psi = \psi^- - \psi^+ = n 2 \upi,
  \label{eq:delta_psi}
\end{equation}
with $n$ being an integer number including zero.
This condition defines a Poincar\'e section in the three dimensional parameter space -- as described in Sect.~\ref{sec:poincare} -- which is used as the basis for further analysis.
The effective dimensionality of the problem is thus reduced to two and the particle motion can be depicted in $\psi^+$-$\mu$-phase space, similar to the previous analysis of the one wave case.
Results from simulations S1 through S4 are shown in Fig.~\ref{fig:4}.

The first setup, S1, is the same as in the previous section.
However, one can formally introduce a second wave with $k^- = -k^+$ and zero amplitude ($\epsilon^- = 0$) to test the analysis using a Poincar\'e section in the three dimensional parameter space.
The result is shown in Fig.~\ref{fig:4}\,a) and looks very similar to Fig.~\ref{fig:3}.
As expected, the data matches well with the lines of $d\psi/d\mu=0$ (black dashed lines) and the positions of the equilibrium points (black crosses) which have been derived in Sect.~\ref{sec:circular_wave}.
However, the phase space trajectories are now represented by a series of discrete points which do not necessarily form unbroken lines.
This is most obvious near $\theta = \upi/2$, where the particles hardly propagate along the background magnetic field and the condition $\Delta\psi = n 2 \upi$, as defined in Eq.~\eqref{eq:delta_psi}, is seldom met.
Thus, data is scarce in this region.

Phase space trajectories change with increasing amplitude $\epsilon^-$, as can be seen in Fig.~\ref{fig:4}\,b) to d).
Regions of chaotic particle motion can be distinguished from smaller islands where ordered particle motion is found.
We will first discuss the islands, where particles seem to describe closed orbits in the Poincar\'e section.

As discussed in Sect.~\ref{sec:poincare}, a stationary point in the Poincar\'e section indicates that particle motion is periodic in the full three-dimensional phase space.
We assume that each island surrounds one such stationary point (see the light blue dot at $\psi^+=\frac{3}{2}\upi$ slightly above the lowest black cross in Fig.~\ref{fig:4}\,b), although we do not have particle data to support the existence of all of them.
The ordered structures around these stationary points represent trajectories which are at least quasi-periodic in three-dimensional phase space.

For a wave with elliptical polarization (S2 and S3, see Fig.~\ref{fig:4}\,b and c) we notice that stationary points (and the corresponding islands) appear close to the positions of the three stable equilibria found in the one wave case (see Figs.~\ref{fig:1} and~\ref{fig:4}\,a, labels 1, 2, 3).
An additional fourth island is formed around $\psi^+ = \frac{1}{2}\upi, \, \theta \sim \frac{3}{4}\upi$, which is not related to one of the previously known equilibrium points.
As the amplitude $\epsilon^-$ is increased (from S2 to S3) the islands around $\psi^+ = \frac{1}{2}\upi$ change their shapes and become smaller, while those around $\psi^+ = \frac{3}{2}\upi$ keep their triangular appearance, but also shrink slightly.
The positions of the stationary points move to larger (smaller) $|\mu|$ for $\psi=\frac{1}{2}\upi$ ($\psi=\frac{3}{2}\upi$).

Finally, for the linearly polarized wave with $\epsilon^- = \epsilon^+$ the shape and position of the ordered islands is symmetric about $\theta = \frac{1}{2}\upi$ (S4, see Fig.~\ref{fig:4}\,d).
Interestingly the islands around $\psi^+ = \upi/2$ are split in two, as can be seen by the accumulation of green dots near $\psi^+=\frac{1}{2}\upi, \, \theta \sim \frac{3}{4}\upi$ or the purple dots around $\psi^+=\frac{1}{2}\upi, \, \theta \sim \frac{1}{4}\upi$, which each correspond to one particle.
We assume that these features are associated to two stationary points in the Poincar\'e section which lie between the two green and purple ellipses, respectively.

In between the different islands of ordered particle orbits it is hard to find a clear structure.
Particles seem to travel through phase space in a random motion, which hints at chaotic behavior.
Panels e) and f) in Fig.~\ref{fig:4} show the same data as panels c) and d), respectively, corresponding to simulations S3 and S4.
In each of these two panels, two individual particles are highlighted in color, and the discrete points denoting the particles' coordinates in phase space at different times are connected via straight lines to give an impression of particle motion in the Poincar\'e section.
It can be clearly seen that these particles do not enter the ordered islands, but travel through the surrounding space only.
At times it is possible that a particle orbits one of the stationary points (see e.g.~the red line in panel f).
However, these orbits are not stable and the particle returns to a chaotic motion.

Changes of the direction of propagation relative to the background magnetic field $\vec{B}_0$ are also possible, although not frequent (see Fig.~\ref{fig:4}\,e and f).
This result is interesting from the point of view of quasi-linear theory \citep[e.g.][]{schlickeiser_1989}:
The standard magnetostatic approach to wave-particle interaction employs the limit of $\epsilon \rightarrow 0$ and finds a resonance gap at $\mu=0$, i.e. no particle scattering across that line.
Our system, however, shows that a resonance gap can be avoided if finite amplitude ($\epsilon > 0$) waves are considered.
Even in the relatively simple case of a linearly polarized wave we find chaotic orbits which enable particles to cross $\mu=0$.

To conclude the discussion of the numerical results it has to be kept in mind that the trajectories presented in Fig.~\ref{fig:4} do not represent a universal picture.
The size and shape of the ordered islands in panels b) through d) depend on the amplitude of the waves.
For the case of the linearly polarized wave (Fig.~\ref{fig:4}\,d) we have carried out additional simulations (not shown here) with varied amplitudes $\epsilon^\pm$.
For smaller amplitudes the region of stochastic particle transport shrinks, as expected \citep{murakami_1982}.
This is also consistent with the idea that the stochastic regions vanish below a threshold amplitude \citep{balakirev_1989}.
For increasing amplitude, on the other hand, the ordered islands shrink and it seems possible that they can vanish completely.
However, for the example discussed here this would require an amplitude which is larger than $B_0$.

\section{Summary and Conclusions}
\label{sec:conclusions}

We have discussed the problem of particle transport in magnetostatic slab fluctuations.
In Sect.~\ref{sec:model} we have laid out the general model of a transverse perturbation on top of a static background magnetic field $\vec{B}_0 = B_0 \vec{e}_z$.
We have derived the Hamiltonian for a charged particle in this general field setup and have discussed the constants of motion.
We have then continued to the equations of motion of the particle in Sect.~\ref{sec:equations_of_motion}, where we made use of a new set of field line coordinates.
These field line coordinates allow us to map the trajectory of the particle to the $z=0$ plane.
Switching to yet another set of variables has allowed us to finally express the dynamics of the system by three differential equations for the pitch-angle cosine $\mu$ (in the unperturbed field), the phase angle $\varphi_G$ in the $z = 0$ plane, and the $z$-coordinate, Eqs.~(\ref{eq:dot_mu} - \ref{eq:dot_z}).

A detailed analysis for the case of a circularly polarized, monochromatic wave was presented in Sect.~\ref{sec:circular_wave}.
We could show that the equations of motion reduce to two autonomous equations, Eqs.~\eqref{eq:finalmu} and~\eqref{eq:finalpsi}.
Thus, we were able to describe the motion of a charged particle in phase space by only two variables, namely $\mu$ and $\psi$, where $\psi$ describes the phase between the wave and the particle in the $z=0$ plane.
We have demonstrated that the particles follow closed orbits in phase space and that, therefore, no chaotic behavior can be found.
The discussion was concluded by the analysis of stable equilibrium points in phase space.

In Sect.~\ref{sec:linear_wave} we have applied our previous approach to the case of a linearly polarized, monochromatic wave.
By representing the linearly polarized wave by a superposition of two circularly polarized waves with opposite helicity, we are able to derive a set of three differential equations for $\mu$, $\psi^+$, and $\psi^-$.
The latter two variables describe the phase between the particle and the two waves with opposite helicity.
However, other than for the case of the circularly polarized wave, it is not possible to separate the variables and solve the equations analytically.
We have then briefly discussed the lack of equilibrium points in three-dimensional phase space and the introduction of Poincar\'e sections for further analysis of the system.

By means of numerical simulations of test particles in a prescribed, static magnetic field, we have revisited particle transport in the case of circularly and linearly polarized, monochromatic waves in Sect.~\ref{sec:simulations}.
We were able to reproduce our analytical results for the former case, and could show that chaotic behavior can be found in the latter case.
To analyze the linearly polarized wave we have resorted to Poincar\'e sections through phase space, in order to map particle positions in the three-dimensional phase space to a two-dimensional plane.
Besides the chaotic behavior, we could also show that islands of ordered particle motion exist in the Poincar\'e section.
Stationary points in the Poincar\'e sections could be found, indicating fully periodic motion in full three-dimensional phase space for these orbits.
These do not, however, correspond to equilibrium points of the full system.
The calculation of the stationary points in the Poincar\'e section requires either the knowledge of the trajectory in three-dimensional phase space (which cannot be obtained analytically, as we have shown) or a complicated perturbation ansatz.

While the problem of charged particle transport in magnetostatic slab mode field fluctuations is by no means new and has been treated recurrently and extensively in the literature, our work presents a novel approach.
By mapping the particle trajectory onto the $z=0$ plane we are able to derive a new and elegant set of equations of motion.
For the case of a monochromatic, circularly polarized wave we can find an exact analytical solution to the problem.
Our solution holds for finite amplitudes of the magnetic perturbations and requires no further approximations.

The derivation of particle trajectories in $\mu$-$\psi$-phase space clearly shows that particle motion is deterministic.
This might seem counter-intuitive as one expects resonant scattering processes for waves fulfilling the cyclotron resonance condition.
However, a single circularly polarized wave only yields closed orbits in phase space and no effective scattering.
Chaotic motion can, therefore, only exist when two or more circularly polarized waves are present.

The numerical test particle simulations presented in Sect.~\ref{sec:simulations} are primarily meant to illustrate our point.
However, the setup for our simulations could also be used as a test case for the validation of numerical codes, such as the \textit{Streamline} code by \cite{dalena_2012}.
The setup is easy to implement and the results can be directly compared to the analytical solutions.

For the monochromatic, circularly polarized wave our simulations show the same closed orbits in phase space as our analytical model predicts.
However, for the case of a linearly polarized wave, where no analytical solution to the equations of motion can be found, these simulations can contribute to a deeper understanding of the problem.
By mapping the phase space trajectories of the test particles to a two-dimensional Poincar\'e section through the full phase space, we could show that individual islands of stable particle orbits exist.
Inside these islands particle motion might be ordered, while it is chaotic elsewhere.
Although no equilibrium points in phase space can be derived analytically, the numerical results suggest that there are stationary points in the Poincar\'e section.
Particle trajectories crossing these points are periodic in three-dimensional phase space.

A more involved study of the chaotic basins and regular orbits in the Poincar\'e sections will be performed in future papers, where we will also extend the study from the polarization to the wave form, i.e., consider two waves with different wave number, and to a number of waves greater than two.
For small amplitudes we expect that islands of ordered motion in phase space will persist in the Poincar\'e sections.
As seen for the case of two plasma waves, the position of the islands may shift and their size may decrease if the amplitudes of the waves are increased.

It might also be worthwhile to study non-slab geometry with more advanced two-dimensional simulations.
However, an analytic approach is not feasible, as the equations of motion include infinite sums of Bessel functions in the case of obliquely propagating waves.
In some cases approximate solutions may be found \citep[see e.g.][]{palmadesso_1972} or the equations of motion can be integrated numerically \citep[e.g.][]{smith_1978}.
It is therefore questionable whether analytic considerations can be of general use, or if numerical test particle simulations might be the better choice straight away.


We acknowledge the use of the \emph{ACRONYM} code and would like to thank the developers (Verein zur F\"orderung kinetischer Plasmasimulationen e.V.) for their support.

Funding:
This work is based upon research supported by the National Research Foundation and Department of Science and Technology.
Any opinion, findings and conclusions or recommendations expressed in this material are those of the authors and therefore the NRF and DST do not accept any liability in regard thereto.

CS would like to thank J\"org B\"uchner for having him as a guest at the MPS and the Max-Planck-Society for granting a stipend.

RV acknowledges the financial support of the Academy of Finland (project 267186).

\appendix

\section{Change of the actual pitch-angle}
\label{app:actual_pitch_angle}

The pitch-angle cosine $\mu$ discussed in Sects.~\ref{sec:model}, \ref{sec:equations_of_motion} and~\ref{sec:circular_wave} is the angle between the background magnetic field and the particle momentum.
This is different from the actual pitch-angle, especially when the magnetic field perturbations are no longer negligible.
For a more detailed analysis of particle motion or for deriving transport theory it is important to keep in mind that the actual pitch angle $\alpha$ behaves differently from expectations based on $\mu$.

We find that
\begin{eqnarray}
  \cos \alpha &=& \frac{\vec{v} \cdot \vec{B}}{|\vec{v}| |\vec{B}|} = \frac{v_x \delta B_x + v_y \delta B_y + v_z B_0}{v\sqrt{\delta B_x^2 + \delta B_y^2 + B_0^2}}\\
  &=& \frac{\dot x \epsilon \cos kz - \dot y \epsilon \sin kz + v\mu}{v\sqrt{1+\epsilon^2}}.
\end{eqnarray}
We want to express $\cos \alpha$ using the variables $\psi, \mu$:
\begin{eqnarray}
  \cos\alpha &=& \frac{(\dot{x}_F+v\mu \epsilon k z)\epsilon \cos kz - (\dot{y}_F-v\mu \epsilon k z)\epsilon \sin kz + v\mu}{v\sqrt{1+\epsilon^2}}\\
  &=& \frac{(\dot{X}_G + v\mu\epsilon\cos kz)\epsilon \cos kz - (\dot{Y}_G - v\mu\epsilon\sin kz)\epsilon \sin kz + v\mu}{v\sqrt{1+\epsilon^2}}\\
  &=& \epsilon\frac{\dot{X}_G\cos kz - \dot{Y}_G\sin kz +v\mu\epsilon(\cos^2 kz +\sin^2 kz)}{v\sqrt{1+\epsilon^2}}+\frac{\mu}{\sqrt{1+\epsilon^2}}\\
  &=& \epsilon \frac{(\dot{R}_G\cos\varphi_G - R_G\dot{\varphi}_G\sin\varphi_G)\cos kz -(\dot{R}_G\sin\varphi_G + R_G\dot{\varphi}_G\cos\varphi_G)\sin kz}{v\sqrt{1+\epsilon^2}}\nonumber\\
  && + \mu\frac{1+\epsilon^2}{\sqrt{1+\epsilon^2}}\\
  &=& \epsilon \frac{\dot{R}_G(\cos\varphi_G \cos kz - \sin\varphi_G \sin kz) -{R}_G\dot{\varphi}_G(\sin\varphi_G \cos kz +\cos\varphi_G \sin kz)}{v\sqrt{1+\epsilon^2}}\nonumber\\
  && + \mu\frac{1+\epsilon^2}{\sqrt{1+\epsilon^2}}\\
  &=& \epsilon\frac{\dot{R}_G\cos(\varphi_G+kz)-R_G\dot{\varphi}_G\sin(\varphi_G+kz)}{v\sqrt{1+\epsilon^2}} + \mu\frac{1+\epsilon^2}{\sqrt{1+\epsilon^2}}\\
  &=& \epsilon\frac{-\epsilon v \mu \cos^2(\varphi_G + kz)- \epsilon v \mu \sin^2(\varphi_G+kz)+\Omega_0 R_G \sin(\varphi_G+kz)}{v \sqrt{1+\epsilon^2}}\nonumber\\
  && + \mu\frac{1+\epsilon^2}{\sqrt{1+\epsilon^2}}\\
  &=& \frac{\mu+\epsilon\sqrt{1-\mu^2}\sin\psi}{\sqrt{1+\epsilon^2}}.
  \label{eq:cos_alpha_psi}
\end{eqnarray}
Thus, for a constant relative phase, we get a solution of constant $\cos \alpha$.
Using $F=\epsilon \sin\psi$ and the expression from Eq.~\eqref{eq:xx}, we get
\begin{eqnarray}
  \cos \alpha &=& \frac{\mu}{\sqrt{1+\epsilon^2}}+\sqrt{\frac{1-\mu^2}{1+\epsilon^2}}F\\
  &=& \frac{2\Omega_0 kv \mu + (kv\mu -\Omega_0)^2-C}{2\Omega_0 k v \sqrt{1+\epsilon^2}}\\
  &=& \frac{k^2v^2\mu^2+\Omega_0^2-C}{2\Omega_0 k v \sqrt{1+\epsilon^2}}.
  \label{eq:cos_alpha}
\end{eqnarray}
This result clearly shows the nonlinear relation of $\cos\alpha$ and $\mu$.

\begin{figure}
  \centering
  \includegraphics[width=0.8\linewidth]{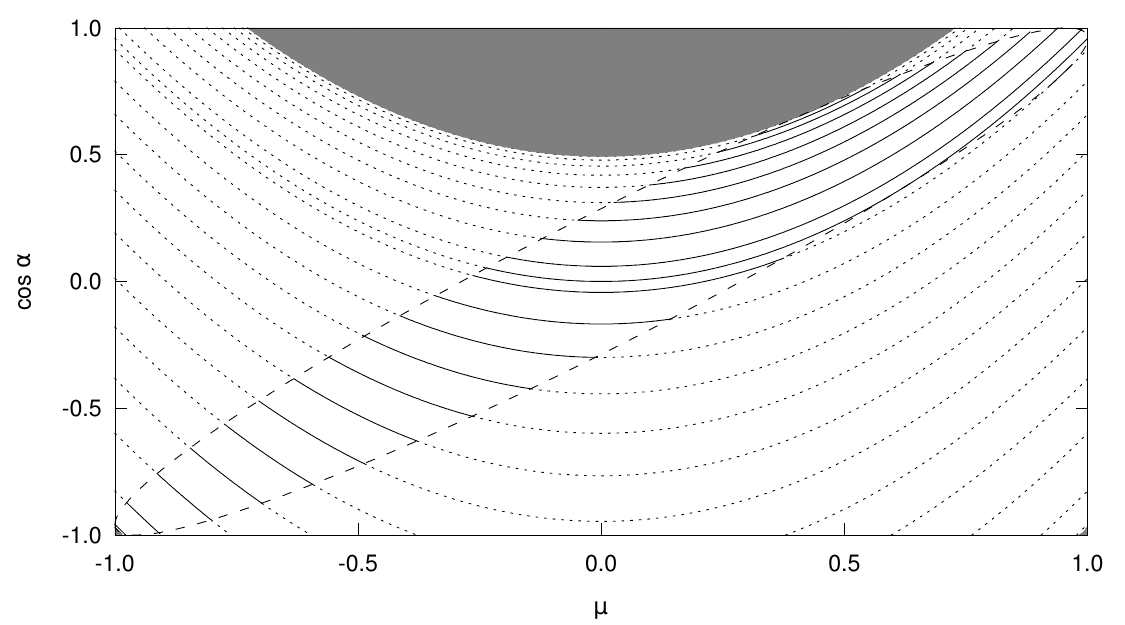}
  \caption{
  Relation of the actual pitch-angle cosine $\cos\alpha$ in the perturbed magnetic field and $\mu$ in the unperturbed field according to Eq.~\eqref{eq:cos_alpha}.
  We show $\cos\alpha$ as a function of $\mu$ for $\kappa=2$, $\epsilon=0.3$ and the same values of $C$ that have been used in Fig.~\ref{fig:1}.
  Physical solutions (black solid lines) can be found in the region between the two dashed boundary lines.
  The continuations of these solutions (black dotted lines) can also be obtained from Eq.~\eqref{eq:cos_alpha}, but are not realized by the particles.
  Note that $C$ is limited (see Eq.~\eqref{eq:const_integration}) and that no solutions can be obtained for $C > C_\mathrm{max}$ (grey areas at the bottom corners) or $C < C_\mathrm{min}$ (grey area at the top).
  }
  \label{fig:5}
\end{figure}

While the relation itself is not very complicated, the behavior of $\cos\alpha$ might not seem intuitive.
We therefore plot the relation of $\cos\alpha$ and $\mu$ in Fig.~\ref{fig:5} according to Eq.~\eqref{eq:cos_alpha}.
We choose the same values for $C$ that have been used to plot the trajectories in Fig.~\ref{fig:1}.
The constant of integration, $C$, can only take values in a limited range, defined by the physical parameters $\kappa$ and $\epsilon$ of the system.
Thus, there are areas in Fig.~\ref{fig:5} where no physical solution to Eq.~\eqref{eq:cos_alpha} can be obtained (marked in grey).

Furthermore it should be noted that the trajectory of any given particle does not cover the whole range $-1 < \mu < 1$ (see Fig.~\ref{fig:1} to recall the shape of the particle orbits in phase space).
We find that the minima and maxima in $\mu$ occur at $\psi = \frac{1}{2} \upi$ or $\psi = \frac{3}{2} \upi$.
Thus, the available range of the actual pitch-angle cosine is given by
\begin{equation}
  \cos \alpha = \frac{\mu \pm \epsilon \sqrt{1-\mu^2}}{\sqrt{1+\epsilon^2}}
\end{equation}
according to Eq.~\eqref{eq:cos_alpha_psi}.
We plot the two solutions as dashed lines in Fig.~\ref{fig:5}.
Solid curves between these boundaries show the physical solutions of Eq.~\eqref{eq:cos_alpha} for the given values of $C$.
The continuations of these curves outside of the boundaries (dotted curves) are not realized by the particles.

\begin{figure}
  \centering
  \includegraphics[width=0.8\linewidth]{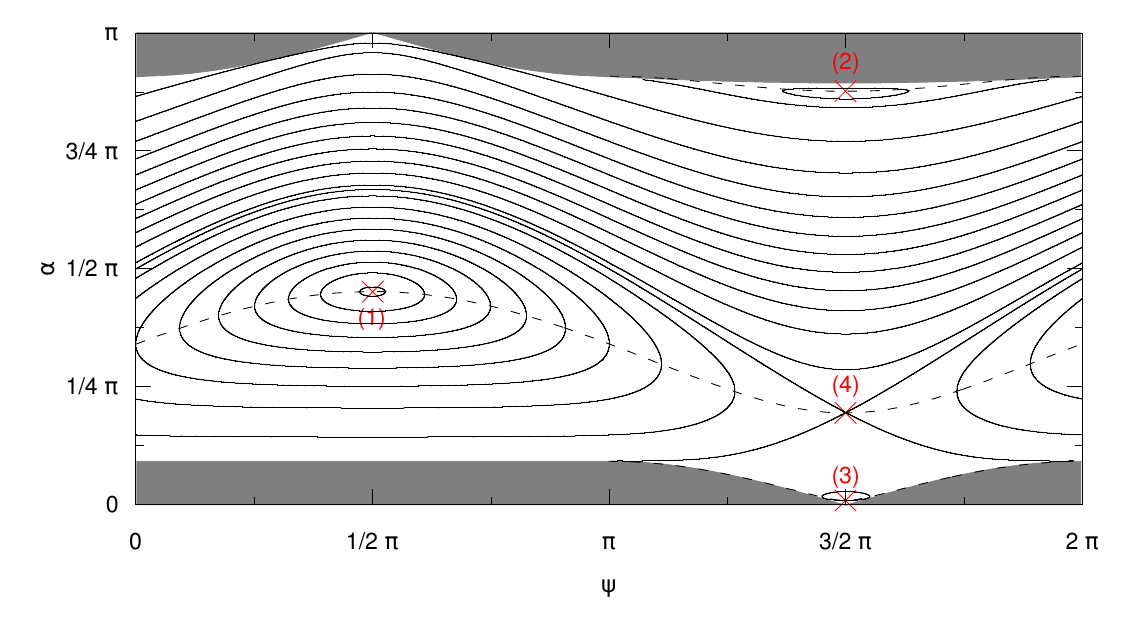}
  \caption{
  Phase space trajectories (solid lines) in $(\psi,\,\alpha)$ coordinates for $\kappa=2$, $\epsilon=0.3$ and the same values of $C$ that have been used in Fig.~\ref{fig:1}.
  The lines connecting the turning points $d\psi/d\alpha=0$ are shown as dashed lines, the equilibrium points are marked in red and labeled according to our findings in Sect.~\ref{sec:circular_wave}.
  Grey areas denote regions in phase space that are not accessible to the particles.
  }
  \label{fig:6}
\end{figure}

Finally, we transform the particle orbits from Fig.~\ref{fig:1} into the coordinate system $(\psi,\,\alpha)$ and show the result in Fig.~\ref{fig:6}.
Mapping $\psi$-$\mu$-phase space to the new coordinate system produces trajectories with slightly different shapes.
The four equilibrium points (marked by the red crosses and labels) and the turning points $d\psi/d\alpha = 0$ (dashed lines) can be recovered.
Note that certain regions in the new phase space are not accessible to the particles (grey shaded areas).

\section{Numerical accuracy of the test particle simulations}
\label{app:num_error}

The question of numerical accuracy and potential errors often arises when numerical simulations are employed to model complex physical processes.
One way to discuss numerical accuracy is to test the consistency of the simulation itself.
For self-consistent methods, such as particle-in-cell (PiC), a variety of scenarios can be analyzed to check for numerical errors \citep[see e.g.][for some typical validation methods]{melzani_2013}.
The ACRONYM PiC code \citep{kilian_2011}, which serves as the basis for the magnetostatic code used in this work, has been thoroughly tested over the years.
During the process new test cases have been established, such as the analysis of wave modes \citep{kilian_2017} and their damping characteristics \citep{schreiner_2017_b}, which can be used to check for inconsistencies in the interplay of particle motion and electromagnetic fields.

For other types of codes, such as test particle codes \citep[e.g.][]{dalena_2012} or the hybrid-magneto-hydrodynamic (hybrid-MHD) approach \citep[e.g.][]{lange_2013}, less rigorous testing methods might be sufficient to ensure correct behavior of test particles.
A simple test could be to simulate gyrating particles in a static magnetic field $\vec{B}_0 = B_0 \, \vec{e}_z$.
The particles' orbits can be analyzed to obtain their Larmor radii.
These in turn can be compared to the exact analytic value to check for numerical errors and to get a measure for numerical accuracy.
\citet{lange_2013} use this method to automatically adjust the numerical time step at the beginning of their test particle MHD simulations.
As a rule of thumb, ten time steps per gyration are typically sufficient to reduce numerical errors to negligible levels.

Basically, the simulation described in Sect.~\ref{sec:simulations_one} can also be used for code validation.
Particle orbits in phase space can be drawn (see Fig.~\ref{fig:3}) and compared to the analytic results (see Fig.~\ref{fig:1}).
The study can be done on a qualitative level by comparing the shapes of the particle orbits in theory and simulation and by checking the positions of equilibrium points.
However, preparing the analytical results might be cumbersome and a qualitative study might not be as convincing as a quantitative error measure.

We therefore suggest a different approach:
Particle orbits in $\psi$-$\mu$-space are lines of constant $C$, as was derived in Eq.~\eqref{eq:const_integration} in Sect.~\ref{sec:circular_wave}.
In the analytic model we found that $C$ is a constant of motion and can thus be calculated at $t=0$.
This also holds in the numerical simulations.

To check for numerical accuracy it is therefore sufficient to calculate $C(t_0)$ for each particle at the beginning of the simulation using its initial $\psi$ and $\mu$.
The calculation of $C$ can be repeated in each consecutive (output) time step $t$, then using $\psi(t)$ and $\mu(t)$.
Deviations from $C(t_0)$ hint at numerical errors or inconsistencies in the code.
Note that this test can be performed without any knowledge of the analytical model presented in Sect.~\ref{sec:circular_wave}.

\begin{figure}
  \centering
  \includegraphics[width=0.8\linewidth]{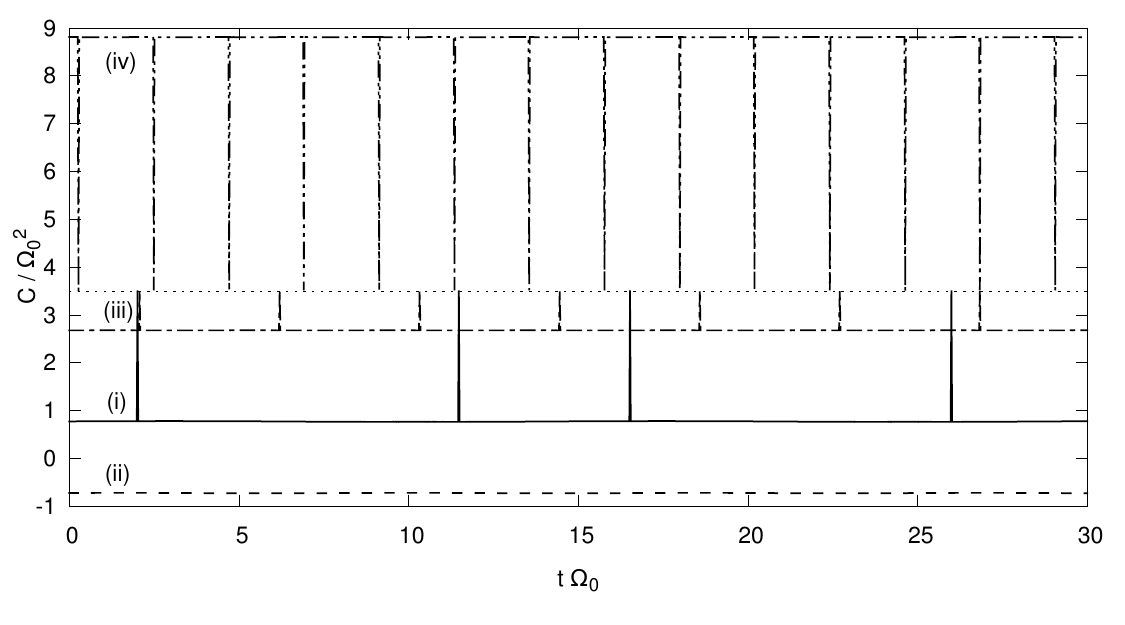}
  \caption{
  The value of the constant of integration, $C$, from Eq.~\eqref{eq:const_integration} is plotted over time for four different particles (solid, dashed, dash-dotted, and dash-dot-dotted lines).
  The orbits of particles (i) through (iv) are marked in orange and labeled with the same Roman numbers in Fig.~\ref{fig:3}.
  Crossings of $\psi=0$ are denoted by vertical lines, where the value of $C$ is deliberately set to $C(\psi\!=\!0)/\Omega_0^2 = 3.5$ (dotted line) for diagnostic purposes.
  }
  \label{fig:7}
\end{figure}

We show results from our simulation in Fig.~\ref{fig:7}.
For four particles (labeled by Roman numbers, see also Fig.~\ref{fig:1}) we track $C$ over the course of the simulation.
As can be seen, the value of $C$ remains constant for all four particles, except for the occurrence of vertical lines in regular patterns.
These vertical lines however do not show numerical errors, but are introduced for diagnostic purposes.
Each time a particle crosses $\psi=0$ the value of $C$ is deliberately set to $C(\psi\!=\!0)/\Omega_0^2 = 3.5$ (see dotted line in Fig.~\ref{fig:7}).
This allows to obtain some more details about particle motion and explains the patterns of the vertical lines:
The trajectory of particle (i) is limited in $\psi$ and crosses $\psi=0$ twice per orbit with the second crossing being in opposite direction compared to the first one.
Particle (ii) is also limited in $\psi$, but does not cross $\psi=0$.
Therefore, no vertical line is obtained in Fig.~\ref{fig:7}.
Finally, particles (iii) and (iv) traverse the entire range of $\psi$, as can be seen in Fig.~\ref{fig:3}, and cross $\psi=0$ once per orbit and each time in the same direction.

\bibliographystyle{jpp}
\bibliography{ref}

\end{document}